\documentclass[11pt]{article}
 
\usepackage{latexsym, amsmath}
\newtheorem{theorem}{Theorem}
\newtheorem{lemma}{Lemma}
\newtheorem{corollary}{Corollary}

\newtheorem{proposition}{Proposition}

\newenvironment{proof}{{\it Proof:\/}}{\hfill $\Box$\\ }

\newcommand{\F}{\mbox{\bf  F}}

\title{On Deciding Deep Holes of Reed-Solomon Codes}
                         
\date{}
                                                                  
\author{Qi Cheng and Elizabeth Murray\thanks{School of Computer Science,
the University of Oklahoma,
Norman, OK 73019, USA.
Email: {\tt \{qcheng, wumpus\}@ou.edu.}
This research is partially supported by NSF Career
Award CCR-0237845
}
}

\begin{document}
                                                                               
\maketitle

\begin{abstract}
For generalized Reed-Solomon codes,
it has been proved \cite{GuruswamiVa05} that the problem of determining if a
received word is a deep hole is co-NP-complete.
The reduction relies on the fact that 
the evaluation set of the code can be exponential 
in the length of the code -- 
a property that practical codes do not usually possess.
In this paper, we first presented a much simpler proof of the same result.
We then consider the problem for standard
Reed-Solomon codes,
i.e. the evaluation set consists of
all the nonzero elements in the  field. 
We reduce the problem of identifying deep holes to 
deciding whether an absolutely irreducible 
hypersurface over a finite field
contains a rational point whose coordinates are pairwise distinct
and nonzero.
By applying Schmidt and Cafure-Matera estimation of rational points
on algebraic varieties, we prove that
the received vector $(f(\alpha))_{\alpha \in \F_q}$ 
for Reed-Solomon $[q,k]_q$, $k < q^{1/7 - \epsilon}$, 
cannot be a deep hole, whenever $f(x)$ is a polynomial
of degree $k+d$ for  $1\leq d  < q^{3/13 -\epsilon}$.

\end{abstract}

\section{Introduction}

A signal, when transfered over a long distance, always has
a possibility of being corrupted.
Error-detecting and error-correcting codes 
make the modern communication possible.
The Reed-Solomon codes are very popular in engineering 
a reliable channel due to their simplicity, burst error-correction
capabilities, and
the powerful decoding algorithms they admit. 

Let $\F_q$ be the finite field with $q$
elements, where $q$ is a prime power.
The encoding process of generalized Reed-Solomon codes
can be thought of as a map from $\F_q^k \rightarrow \F_q^n$, in which
a message $(a_1, a_2, \cdots, a_k)$ is mapped to a vector
 $$ (f(x_1), f(x_2), \cdots, f(x_n)),    $$
where $f(x) = a_k x^{k-1} + a_{k-1} x^{k-2} + \cdots + a_1 \in \F_q[x]$
and $\{x_1, x_2, \cdots, x_n \} \subseteq \F_q$ is
called the evaluation set. (Note that different encoding schemes are
possible.)

It is not difficult to see that the set of codewords formed in this
manner is a linear subspace of $\F_q^n$ which has dimension $k$. 
Reed-Solomon codes are therefore {\em linear codes}, because they 
are linear subspaces of $\F_q^n$, where $n$ is the length of a codeword.

The {\em Hamming distance} between two codewords is the number of coordinates
in which they differ -- or one can think of this as the number of
modifications required to transform one vector into another. A 
{\em Hamming ball of radius $m$} is
a set of vectors within Hamming distance $m$ to some vector in $\F_q^n$. 
The minimum distance of a code is the smallest
distance between any two distinct codewords, and is a measure of
how many errors the code can correct or detect. 
The {\em covering radius of a code} is the maximum possible distance
from any vector in $\F_q^n$ to the closest codeword.
A deep hole is a vector which achieves this maximum.
The minimum distance of Reed-Solomon codes is $n-k +1$.
The covering radius of Reed-Solomon codes is $n-k$.

A code is useless without a {\em decoding algorithm}, which takes some
{\em received word} (a vector in $\F_q^n$) 
and outputs a message. The message should correspond,
ideally, 
to the codeword which is closest, with respect to Hamming distance, to
the received word. If we assume that each coordinate in a received word is
equally likely to be in error, then the closest codeword is the most likely
to be the intended transmission.  

Standard Reed-Solomon codes use $\F_q^*$ as
their evaluation set. 
If the evaluation set is $\F_q$,
then the code is called an {\em extended Reed-Solomon code}.
If the evaluation set is the set of rational points in a projective line
over $\F_q$,
then the code is known as a {\em doubly extended Reed-Solomon code}.
In this paper,  we consider standard
Reed-Solomon codes, all of our results can be easily generalized 
to extended Reed-Solomon codes.
The difference between
standard Reed-Solomon codes, extended Reed-Solomon codes and doubly
extended Reed-Solomon codes is not practically significant, but generalized
Reed-Solomon codes are quite unique, as
the evaluation set can be exponentially larger than
the length of a codeword.

\subsection{Related Work}
The pursuit of efficient decoding algorithms 
for Reed-Solomon codes has yielded intriguing results.
If the radius of a Hamming ball centered at some 
received word is less than half the
minimum distance, there can be at most one codeword in the
Hamming ball. 
Finding this codeword is called {\em unambiguous
decoding}. It can be efficiently solved, see \cite{BerlekampWe86}
for a simple algorithm.

If the radius is less than $n - \sqrt{n(k-1)}$, the
problem can be solved by the Guruswami-Sudan algorithm
\cite{GuruswamiSu99}, which
outputs all the codewords inside a Hamming ball. 
If the radius is stretched further, the number of
codewords in a Hamming ball may be exponential.
We then study {\em the bounded distance decoding problem},
which outputs just one codeword in any Hamming ball of a certain radius.
More importantly, we can remove the restriction on radius and
investigate the {\em maximum likelihood decoding 
problem}, which is the problem of computing the closest 
codeword to any given vector in $\F_q^n$.

The question on decodability of Reed-Solomon codes
has attracted attention recently, due to
recent discoveries on the relationship between
decoding Reed-Solomon codes and 
some number theoretical problems.
Allowing exponential alphabets, Guruswami and Vardy 
proved  that the maximum likelihood decoding is
NP-complete. They essentially showed 
that deciding deep holes is co-NP-complete.
When the evaluation set is precisely the whole field or $\F_q^*$,
an NP-completeness result is hard to obtain,
Cheng and Wan \cite{ChengWa04} managed to prove that decoding problem
of  Reed-Solomon codes at certain radius is
at least as hard as the discrete logarithm problem
over finite fields. In this paper, we wish to establish
an additional connection between decoding 
of standard Reed-Solomon codes and a classical number-theoretic problem 
-- that of determining the number of rational points on an algebraic 
hypersurface. 

\subsection{Our Results}

The decoding problem of Reed-Solomon codes can be reformulated
into the problem of {\em curve fitting} or {\em noisy polynomial
reconstruction}.  In this problem, we are given $n$ points
$$(x_1, y_1), (x_2, y_2), \cdots, (x_n, y_n)$$
in $\F_q^2$.  The goal is to find polynomials of degree $k-1$ that
pass as many of the $n$ points as possible.
Note that all the $x$-coordinates are distinct.

Given the received word $w = (y_1, y_2, \cdots, y_n)$, 
we are particularly interested in
the polynomial obtained by  interpolating the $n$ points.
\begin{eqnarray*}
  w(x) &=& y_1 { (x-x_2) (x-x_3) \cdots (x-x_n) \over
(x_1-x_2) (x_1-x_3) \cdots (x_1-x_n)  }\\
&&   + \cdots + 
y_i { (x-x_1) \cdots (x-x_{i-1}) (x-x_{i+1})\cdots (x-x_n) \over
(x_i-x_1) \cdots (x_i-x_{i-1}) (x_i-x_{i+1})  \cdots (x_1-x_n)  }  \\
&&
+ \cdots + y_n { (x-x_1) (x-x_2) \cdots (x-x_{n-1}) \over
(x_n-x_1) (x_n-x_2) \cdots (x_n-x_{n-1})  }. 
\end{eqnarray*}
In this paper, we say that a polynomial $w(x)$ {\em generates}
a vector $w \in \F_q^n $ if $w = (w(x_1), w(x_2), \cdots, w(x_n))$.
If the polynomial $w(x)$ has degree $k-1$ or less, $w$ must be a 
codeword, and vice versa 
(since codewords consist of the encodings of all messages 
of length $k$).
If $w(x)$ has degree $k$, $w$ must be a deep hole (as we will later show).
What if it has degree larger than $k$? Can it be a deep hole?

In this paper, we try to answer this question.
If a received word is a deephole,
there is no codeword which is at distance $n-k-1$ or closer
to the received word.
Hence if the distance bound is $n-k-1$, a decoding algorithm
can tell a received word is deephole or not by checking
whether there is a codeword in the Hamming ball of radius $n-k-1$.
This shows that maximum likelihood decoding of Reed-Solomon codes,
as well as the bounded distance decoding
at radius $n-k-1$,
is at least as hard as deciding deepholes.
Observe that the bounded distance
decoding at a distance of $n-k$ or more can be done efficiently.
It is hoped that we can decrease the radius
until we reach the domain of hard problems.

We are mainly concerned with the case
when the evaluation set consists of nonzero elements of the field.  
Notice that for generalized Reed-Solomon code, the bounded
distance decoding at distance $n-k-1$ is NP-hard.
We reduce the problem to deciding whether an absolutely irreducible
hypersurface contains a rational point whose coordinates are pairwise
distinct and nonzero. From the reduction,
we show if $k$ and the degree of $w(x)$
are small, $w(x)$ cannot generate deep holes.
More precisely

\begin{theorem}
Let $q$ be a prime power and $ 1 < k < q^{1/7 - \epsilon}$ 
be a positive integer. 
The vector $(w(\alpha))_{\alpha\in \F_q}$
is not deep hole in Reed-Solomon code $[q,k]_q$
if the degree of $w(x)$ is greater than $k$ but less than 
$k + q^{3/13 - \epsilon}$.
\end{theorem}

Roughly speaking, the theorem indicates that
a vector generated by a low degree polynomial
can not be a deephole, even though it is very far away from
any codeword.

To prove the theorem, we need to estimate
 the number of rational points on an algebraic hypersurface
over a finite field. This problem is 
one of the central problems in algebraic geometry and
finite field theory. Weil, through his proof of
the Riemann Hypothesis for function fields, provided a bound for
 the number of points on algebraic curves. This bound was later 
generalized by Weil and Lang
to algebraic varieties. Schmidt \cite{Schmidt76} obtained
some better bounds for absolutely irreducible hypersurfaces
by elementary means.
In this paper, we will use his results and an 
improved bound, obtained by Cafure and Matera \cite{CafureMa04}
very recently.
But first we give a new proof 
that deciding whether or not a received word is a deep hole is co-NP-complete.
Our reduction is straight-forward and much simpler
that the one constructed by Guruswami and Vardy.


\section{A simple proof that the maximum likelihood decoding
is NP-complete}

We  reduce the following finite field subset sum problem   to 
deep hole problem of generalized Reed-Solomon codes.

\begin{description}
\item[Instance:] A set of $n$  elements
$A = \{x_1, x_2, x_3, \cdots, x_n\} \subseteq \F_{2^m}$, 
an element $b \in \F_{2^m}$ and
a positive integer $k < n$.
\item[Question:] Is there a nonempty subset $\{ x_{i_1}, x_{i_2}, \cdots,
x_{i_k} \} \subseteq A$ of cardinality $k$ such that
$$ x_{i_1} + x_{i_2} + \cdots + x_{i_k} = b. $$
\end{description}

Now consider the generalized Reed-Solomon code $[n,k]_{2^m}$
with evaluation set $A$. Suppose we have a received word 
 $$ w = (f(x_1), f(x_2), \cdots, f(x_n))  $$
where  $f(x) = x^{k+1} + b x^k$. 
If the word is not a deep hole, it is at most $n-k-1$ away from
a codeword. In the other words, there is a polynomial 
$t(x)$ of degree $k-1$ or less such that $f(x) + t(x)$
has at least $k+1$ distinct roots in $A$. Since  $f(x) + t(x)$
is a monic polynomial with degree $k+1$, we have
$$ f(x) + t(x) = x^{k+1} + b x^k + t(x) 
   = (x-x_{i_1})(x- x_{i_2})  \cdots(x - x_{i_{k+1}}),   $$
for some $x_{i_1}, x_{i_2}, \cdots,
x_{i_k}$ in $A$.
Therefore $ x_{i_1} + x_{i_2} + \cdots + x_{i_{k+1}} = b. $

On the other hand, if $ x_{i_1} + x_{i_2} + \cdots + x_{i_{k+1}} = b,$
$f(x) -(x-x_{i_1})(x- x_{i_2})  \cdots(x - x_{i_{k+1}}) $
generates a codeword. It shares $k+1$ values with $w$, thus
has distance less than $n-k-1$ away from a codeword,
so it cannot be a deep hole.

In summary, $w$ is not a deep hole if and only if the answer to
the instance of the finite field subset sum problem is ``Yes''.
Hence the deep hole problem is co-NP-complete.

By a similar argument, we know that
the polynomials of degree $k$ must generate deep holes. Hence

\begin{corollary}
For a generalized Reed-Solomon code $[n,k]_q$,
there are at least $(q-1)q^k$ many deep holes.
\end{corollary}

We remark that the argument cannot work for small evaluation sets,
because the subset sum is easy in that case.
Indeed, if the evaluation set is the whole field and $k>1$,
then a polynomial of degree $k+1$ cannot generate a deep hole.

\section{A hypersurface related to deep holes}

The above argument motivates us to consider  
vectors generated by polynomials of larger degree.
We are given some received word $w$ and we want to know whether or not it
is a deep hole. The received word is generated by $w(x)$ 
of the form $f(x) + t(x)$ where
$f(x)$ is some polynomial containing no terms of degree smaller than
$k$, 
and $t(x)$ is some polynomial containing only terms of degree
$k-1$ or smaller. For purposes of determining whether or not $w$ is a 
deep hole, we fix a monic $f(x)$ 
$$ f(x) = x^{k+d} + f_{d-1} x^{k+d-1} + \cdots + f_0 x^k \in \F_q [x]$$
and let $t(x)$ vary and ask whether
 $f(x) + t(x)$ has $k+1$ roots, or perhaps more.

In its essence, the question is one of finding a polynomial whose leading
terms are $f(x)$, and which has as many zeroes as possible over a certain
field. (As stated earlier, for $k>1$, if $f(x)$ has degree $k$, 
then $w$ is a deep hole.
If $f(x)$ has degree $k+1$, then it is not a deep hole.)  

The most obvious way to approach this problem is to symbolically divide 
$f(x)$ by a polynomial that is the product of $k+1$ distinct (but unknown) 
linear factors, and determine whether or not it is possible to set the 
the leading term of the remainder, i.e., the coefficient
of $x^k$, equal to zero. If the leading coefficient
is $0$, the remainder has degree $k-1$ or less in $x$, which
generates a codeword. The distance between the codeword
and $w$ will be at most $n-k-1$.

A polynomial that is the product of $k+1$ distinct
linear factors will have the elementary symmetric polynomials as coefficients,
$$\Pi = (x-x_1)(x-x_2)...(x-x_{k+1}) = 
x^{k+1} + \pi_1x^k + \pi_2x^{k-1} + ... + \pi_{k+1},$$
where $\pi_i$ is the $i$-th symmetric polynomial in $x_1, x_2, 
\cdots, x_{k+1}$.

Since $\Pi$ is monic, the remainder of $f(x)$ dividing $\Pi$
will be a polynomial in 
$\F_q[x_1, x_2, \cdots, x_{k+1}] [x]$
of degree less than $k+1$. Denote the leading coefficient 
of the remainder by 
$L_{f_0, f_1, \cdots, f_{d-1}}(x_1, x_2, \cdots, x_{k+1})$. 
This is a multivariate polynomial of degree $d$.

As an example, imagine dividing the polynomial
$x^{k+1}$ by $\Pi$. We can easily verify that the leading term of the 
remainder is $-\pi_1 x^{k}$. 
Since we can always find $k+1$ distinct values that will satisfy $\pi_1 = 0$,
we know that $x^{k+1}$ cannot be a deep hole. But, in most cases $w(x)$
will have larger degree and contain many terms, and the remainder will be
a more complex multivariate polynomial in $k+1$ variables, rather than a 
linear polynomial in
$k+1$ variables. If the leading coefficient itself  have a solution where
all roots are distinct and invertible, 
then $f(x) + t(x)$ cannot generate a deep hole.

We now argue that the leading coefficient of the remainder is absolutely
irreducible. We write
$$  L_{f_0, f_1, \cdots, f_{d-1}}(x_1, x_2, \cdots, x_{k+1}) = F_d 
+ F_{d-1} + \cdots + F_0, $$
where $F_i$ is a form  containing all the terms of degree $i$ in $L$.
The polynomial
$L_{f_0, f_1, \cdots, f_{d-1}}(x_1, x_2, \cdots, x_{k+1})$
is absolutely irreducible if $F_d$
is absolutely irreducible. To see this, 
suppose that $L$ can be factored as $L'L''$. Let $F'_{d_1}$
be the form of highest degree in $L'$ and $F''_{d_2}$
be the form of highest degree in $L''$. Then we have
$F_d =F'_{d_1} F''_{d_2}$, a contradiction to the condition that
$F_d$ is absolutely irreducible.

Fortunately $F_d$ 
does not 
depend on $f_i$'s.
\begin{lemma}
The form of the highest degree 
in $L_{f_0, f_1, \cdots, f_{c-1}}(x_1, x_2, \cdots, x_{k+1})$
is exactly 
$L_{0, 0, \cdots, 0}(x_1, x_2, \cdots, x_{k+1})$.
\end{lemma}

\begin{proof}
It can be proved by mathematical induction on $c$.
\end{proof}


In the next section, we argue that the 
term of highest degree in the leading coefficient,
which we will call
$\chi_d (x_1,x_2,...x_{k+1})$, 
is absolutely irreducible. We will actually show that
$\chi_d (x_1,x_2,1,0,0...0)$ is absolutely irreducible.
This is because that $\chi_d (x_1,x_2,1,0,0...0) $ has the same degree as
$\chi_d (x_1,x_2,...x_{k+1})$, if the former is irreducible,
so is the latter.


\begin{lemma}
$ \chi_d (x_1,x_2,1,0,0...0) = \Sigma_{i+j \leq d} {x_1}^i{x_2}^j$.
\end{lemma}

\begin{proof}
We need to compute the leading coefficient of the remainder
after dividing $x^{k+d}$ by $(x-x_1)(x-x_2) (x-1) x^{k-2}$.
It is as same as the leading coefficient of
the remainder after dividing $x^{d+2}$ by $(x-x_1)(x-x_2) (x-1)$.
The remainder is a quadratic polynomial in $x$. When we evaluate it 
at $x_1$, it takes value $x_1^{d+2}$.
When we evaluate it 
at $x_2$, it takes value $x_2^{d+2}$.
When we evaluate it 
at $1$, it takes value $1$. 
By interpolating, we obtain the unique polynomial satisfying 
these conditions. It is

$$  x_1^{d+2} {(x-x_2)(x-1) \over (x_1 - x_2 )(x_1 - 1)} 
+ x_2^{d+2} {(x-x_1)(x-1) \over (x_2-x_1)(x_2-1)} 
+ {(x-x_1)(x-x_2) \over (1-x_1)(1-x_2) }    $$

The leading coefficient is

$$ {x_1^{d+2}  \over (x_1 - x_2 )(x_1 - 1)} 
+ {x_2^{d+2}  \over (x_2-x_1)(x_2-1)} 
+ {1 \over (1-x_1)(1-x_2) },  $$ 
which is equal to $\Sigma_{i+j \leq d} {x_1}^i{x_2}^j$.

\end{proof}

\section{A smooth curve}

The section is devoted to the proof of the irreducibility of
the bivariate polynomial $\Sigma_{i+j \leq d} x^iy^j$.

\begin{lemma}
The curve $f(x, y) = \Sigma_{i+j \leq d} x^iy^j$ 
is absolutely irreducible.
\end{lemma}

\begin{proof}
To show that 
$f(x,y) = \Sigma_{i+j \leq d} x^iy^j$
is absolutely irreducible, we actually prove a stronger
statement that $f(x,y) = 0 $ is a smooth algebraic curve.
It can be verified by simple calculation that
places on the curve at infinity are nonsingular.
Hence  it is sufficient to show all the affine places
on the curve are nonsingular, i.e. that
the system of equations 
$$
\left\{ 
\begin{array}{r@{=}l}
f(x,y)&0\\
\frac{\partial f}{\partial x} & 0\\
\frac{\partial f}{\partial y} & 0 
\end{array}
\right.
$$
has no solution. 

First, it is convenient to write $f(x,y)$ as

$$x^d + (y+1)x^{d-1} + (y^2 + y + 1)x^{d-2} + ... + (y^d + ... + 1).$$ 

We  write $\frac{\partial f}{\partial x}$  as:

$$dx^{d-1} + (d-1)(y+1)x^{d-2} + ...+(y^{d-1} + ... 1)$$ 

and
$\frac{\partial f}{\partial y}$ as 

$$dy^{d-1} + (d-1)(x+1)y^{d-2} + ...+(x^{d-1} + ... 1)$$

%


Assume that there is a solution $(x,y)$ to the system of
equations. Compute 
$(x-y) \frac{\partial f}{\partial x}$:

\begin{eqnarray*}
x \frac{\partial f}{\partial x} &=& dx^{d} 
+ (d-1)(y+1)x^{d-1} + ...+x(y^{d-1} + ... 1)\\
&=& (d+1)x^d + d(y+1)x^{d-1} + \cdots + (y^d + y^{d-1} + \cdots + 1) - f(x,y)\\
y \frac{\partial f}{\partial x} &=& dyx^{d-1} 
+ (d-1)(y^2+y)x^{d-2} + ...+(y^{d} + ... 1)
\end{eqnarray*}

Their difference is:

\begin{eqnarray*}
(x-y) \frac{\partial f}{\partial x} &=& dx^d + [d-(y+1)]x^{d-1} 
+  ... - (y^d + ... + 1)\\
&=& (d+1)x^d + dx^{d-1} + (d-1)x^{d-2} + ... + 1 - f(x,y).
\end{eqnarray*}

Since $f(x,y)$ must be zero,  we know that:

$$(d+1)x^d + dx^{d-1} + (d-1)x^{d-2} + .. + 1 = 0$$

we multiply by the above $x$, and then subtract the original expression to get:

\begin{equation} \label{eqx}
(d+1)x^{d+1} = x^d + x^{d-1} + ... + 1. 
\end{equation}

Repeat the process on $\partial f \over \partial y $, we get

\begin{equation} \label{eqy}
(d+1)y^{d+1} = y^d + y^{d-1} + ... + 1.
\end{equation}

{\em This shows that neither $x$ nor $y$ can be zero.}
Now, observe that $$(x-y)f(x,y) = x^{d+1} + x^d + \cdots + 1 - y^{d+1}
-y^d - \cdots - 1 = 0. $$ 
This means that $(d+2)x^{d+1} = (d+2)y^{d+1}$.
We also know that the right hand sides of \ref{eqx} and \ref{eqy} are
actually 
$\frac{x^{d+1}-1}{x-1}$, and
$\frac{y^{d+1}-1}{y-1}$. So multiplying both sides by $x-1$ 
for \ref{eqx}
and by $y-1$ for \ref{eqy}, we obtain
$$(d+1)x^{d+2} - (d+2)x^{d+1} = 1$$
$$(d+1)y^{d+2} - (d+2)y^{d+1} = 1.$$ 
Hence we have $(d+1)x^{d+2} = (d+1)y^{d+2}$.

If $d+1=0$, we  have $x^d + x^{d-1} + \cdots + 1 = 0$
and 
$ d x^{d-1} + (d-1) x^{d-2} + \cdots + 1 =0,    $
which is as same as $$ {\mathrm{d} \over \mathrm{d}x}  
(x^d + x^{d-1} + \cdots + 1 ) = 0. $$
This means that 
the equation $ x^d + x^{d-1} + \cdots + 1 = 0 $ has a multiple root.
Contradiction.

If $d+2=0$, we conclude that $ x^{d+1} + x^d + \cdots + 1 =0  $
has a multiple root. Again a contradiction.

{\em Now we can assume that
the characteristic of the field does not divide  either $d+1$
or $d+2$.} In particular, this  means that the field must have
odd characteristic.
We have $x^{d+2} = y^{d+2}$ and $x^{d+1} = y^{d+1}$. 
Therefore $x=y$.


In this case, 
$\frac{\partial f}{\partial x}$ 
and
$\frac{\partial f}{\partial y}$ 
 are the same. 
We know that $f(x,y)$ can be
written (since $x=y$) as 
$$(d+1)x^d + dx^{d-1} + (d-1)x^{d-2} + ... + 1 = 0.$$
If $x=y$, then 
$$\frac{\partial f}{\partial x} = (d + (d-1) + ... + 1)x^{d-1} + ... 
((d-1) + (d-2) + ....)x^{d-2} + ... + 1 = 0. $$ 

If we subtract $f(x,y)$ from $\frac{\partial f}{\partial x}$,
we get
$$ (d+1)x^d - ( (d-1) + (d-2) + \cdots + 1) x^{d-1} - 
\cdots - x = 0.    $$ 
Divide the result by $x$,  we have 
$$ (d+1) x^{d-1} - ( (d-1) + (d-2) + \cdots + 1) x^{d-2} - 
\cdots - 1 =  (d+1)x^{d-1} + ( d+ (d-1) + \cdots + 1) x^{d-1} = 0.    $$

This means that ${(d+1)(d+2) \over 2 } x^{d-1} = 0$, 
hence $x=0$, and this is a contradiction.

\end{proof}

\section{Estimation of rational points}

Cafure and Matera \cite[Theorem 5.2]{CafureMa04} 
have obtained the following estimation
of number of rational points on an absolutely irreducible
hypersurface:

\begin{proposition}
An absolutely irreducible $\F_q$-hypersurface in $\F_q^n$
contains at least 
 $$ q^{n-1} - (d-1)(d-2) q^{n - 3/2} - 5d^{13/3} q^{n-2}  $$
many $\F_q$-rational points.

\end{proposition}

We also use the following proposition, proved by Schmidt \cite{Schmidt76}.

\begin{proposition}
Suppose $f_1 (x_1, x_2, \cdots, x_n)$ and $ f_2(x_1, x_2, \cdots, x_n) $
are polynomials of degree not greater than $d$, and they donot have
a common factor. Then the number of $\F_q$-rational solutions
of
$$ f_1 = f_2 = 0$$
is at most $2nd^3 q^{n-2}$.
\end{proposition}

We seek solutions of $L (x_1, x_2,\cdots, x_{k+1}) $ but not of
$$ \Pi_{1\leq i\leq k+1} x_i \Pi_{1\leq i< j\leq k+1} (x_i - x_j) =0.  $$
We count the number of  rational solutions of 
$L (x_1, x_2,\cdots, x_{k+1}) $,
minus the number of rational solutions of
$$
\left\{ 
\begin{array}{r@{=}l}
L (x_1, x_2,\cdots, x_{k+1}) &0\\
\Pi_{1\leq i\leq k+1} x_i \Pi_{1\leq i< j\leq k+1} (x_i - x_j)
& 0
\end{array}
\right.
$$
The number is greater than
$$ q^{k} - (d-1)(d-2) q^{k - 1/2} - 5d^{13/3} q^{k-1}  - 2 (k+1)
[\max(d, {k^2 + k + 2 \over 2})]^3 q^{k-1},   $$
which is greater than $0$ if
$d < q^{3/13 - \epsilon}$ and $ k < q^{1/7 - \epsilon}$.
This concludes the proof of the main theorem.

\section{Concluding Remarks}

It has been proved that for generalized Reed-Solomon codes,
the bounded distance decoding at radius $n-k-1$ is NP-hard.
In this paper, we try to determine the complexity of this problem
for standard Reed-Solomon codes.
We reduce the problem to 
a problem of determining whether a hypersurface contains
a rational point of distinct  coordinates. 
While we didnot solve the problem
completely, we show that for small $k$,
this problem is easy if a vector is generated
by small degree polynomial.
In essential, we ask whether there
exists a polynomial with many distinct
rational roots under the restriction that some coefficients are
prefixed.
This problem bears an interesting comparison
with the active researches \cite{Cohen05} on construction 
of irreducible
polynomial with some prefixed coefficients.


To  solve the problem for every $k$ and every vector,
there are two apparent approaches: 1) Find a better estimation of
number of rational points on the hypersurfaces.
2) Explore the specialty of the hypersurfaces.
From an average argument, it is attempting to conjecture 
that the vectors generated by polynomials of degree
$k$ are the only deep holes possible.
If so, we can completely classify deep holes.
We leave it as a open problem.

%


\begin{thebibliography}{1}

\bibitem{BerlekampWe86}
E.~Berlekamp and L.~Welch.
\newblock Error correction of algebraic block codes.
\newblock U.S. Patent Number 4633470, 1986.

\bibitem{CafureMa04}
A.~Cafure and G.~Matera.
\newblock Improved explicit estimates on the number of solutions of equations
  over a finite field.
\newblock http://www.arxiv.org/abs/math.NT/0405302, 2004.

\bibitem{ChengWa04}
Qi~Cheng and Daqing Wan.
\newblock On the list and bounded distance decodibility of the {Reed-Solomon}
  codes (extended abstract).
\newblock In {\em Proc.\ $45$th IEEE Symp.\ on Foundations of Comp.\ Science},
  pages 335--341, 2004.

\bibitem{Cohen05}
Stephen~D. Cohen.
\newblock Explicit theorems on generator polynomials.
\newblock {\em Finite Fields and Their Applications}, 2005.
\newblock To appear.

\bibitem{GuruswamiSu99}
Venkatesan Guruswami and Madhu Sudan.
\newblock Improved decoding of {Reed-Solomon} and algebraic-geometry codes.
\newblock {\em IEEE Transactions on Information Theory}, 45(6):1757--1767,
  1999.

\bibitem{GuruswamiVa05}
Venkatesan Guruswami and Alexander Vardy.
\newblock Maximum-likelihood decoding of reed-solomon codes is {NP}-hard.
\newblock In {\em Proceeding of SODA}, 2005.

\bibitem{Schmidt76}
W.~Schmidt.
\newblock {\em Equations over Finite Fields. An Elementary Approach}, volume
  536 of {\em Lecture Notes in Mathematics}.
\newblock Springer-Verlag, 1976.

\end{thebibliography}
\end{document}